\documentclass[preprint,5p,authoryear]{elsarticle}
\usepackage{amssymb}
\usepackage{tabularx}
\usepackage{makecell}
\usepackage{rotating}
\usepackage{threeparttable}
\usepackage{booktabs}
\usepackage{array}
\usepackage{multirow}

\usepackage{subfigure}
\usepackage{verbatim}
\usepackage{lineno,hyperref}
\modulolinenumbers[5]
\usepackage{amsmath,graphicx}
\usepackage{fixltx2e}
\usepackage{amsmath,bm}
\usepackage{color}
\usepackage{xcolor}
\usepackage[varg]{txfonts}
\usepackage{siunitx}
\usepackage{ulem}

\PassOptionsToPackage{numbers,sort&compress}{natbib} 
\renewcommand\cite[1]{\citeauthor{#1}, \citeyear{#1}}
\usepackage{arydshln} 

\journal{Speech Communication, * is Corresponding author, E-mail addresses: longbiao\_wang@tju.edu.cn (L. Wang)}

\begin{document}

\begin{frontmatter}

\title{LORT: Locally Refined Convolution and Taylor Transformer for Monaural Speech Enhancement}
\author{Junyu Wang$^{1}$, Zizhen Lin$^{2}$, Tianrui Wang$^1$, Meng Ge$^1$, Longbiao Wang$^{1,3,*}$, Jianwu Dang$^4$}

\address{$^1$Tianjin Key Laboratory of Cognitive Computing and Application, College of Intelligence and Computing, Tianjin University, Tianjin, China\\ 
    $^2$School of Electronic Information, Sichuan University, Sichuan, China\\ $^3$Huiyan Technology (Tianjin) Co., Ltd, Tianjin, China\\
    $^4$Shenzhen Institute of Advanced Technology, Chinese Academy of Sciences, China}

\begin{abstract}
Achieving superior enhancement performance while maintaining a low parameter count and computational complexity remains a challenge in the field of speech enhancement. In this paper, we introduce LORT, a novel architecture that integrates spatial-channel enhanced Taylor Transformer and locally refined convolution for efficient and robust speech enhancement. We propose a Taylor multi-head self-attention (T-MSA) module enhanced with spatial-channel enhancement attention (SCEA), designed to facilitate inter-channel information exchange and alleviate the spatial attention limitations inherent in Taylor-based Transformers. To complement global modeling, we further present a locally refined convolution (LRC) block that integrates convolutional feed-forward layers, time-frequency dense local convolutions, and gated units to capture fine-grained local details. Built upon a U-Net-like encoder-decoder structure with only 16 output channels in the encoder, LORT processes noisy inputs through multi-resolution T-MSA modules using alternating downsampling and upsampling operations. The enhanced magnitude and phase spectra are decoded independently and optimized through a composite loss function that jointly considers magnitude, complex, phase, discriminator, and consistency objectives. Experimental results on the VCTK+DEMAND and DNS Challenge datasets demonstrate that LORT achieves competitive or superior performance to state-of-the-art (SOTA) models with only 0.96M parameters, highlighting its effectiveness for real-world speech enhancement applications with limited computational resources.
\end{abstract}
 

\end{frontmatter}

\section{Introduction}

Speech enhancement is a key technology in speech processing, designed to improve the clarity and intelligibility of speech signals in noisy environments. Its applications span critical domains such as telecommunication systems, hearing aids, automatic speech recognition (ASR), and multimedia production. By separating target speech from environmental noise, speech enhancement ensures reliable communication in challenging scenarios, including aircraft cockpits, emergency response systems, and smart home devices. 

The core objective of speech enhancement is to recover clean speech from noisy observations while preserving its spectral-temporal characteristics. Generally, this process can be broadly categorized into time-domain and time-frequency (T-F) domain approaches. Time-domain speech enhancement use techniques such as adaptive filtering to directly process the raw speech signal \citep{SEGAN, convtasnet, ARFDCN}. Advantages include computational efficiency and phase preservation, but can be limited in non-stationary noise environments due to fixed filter coefficients. T-F domain methods convert the signal into T-F representations by short-time Fourier transform (STFT) to analyze the magnitude and phase components separately \citep{MetricGAN, phasen}. This approach achieves noise suppression by exploiting frequency characteristics, but introduces the challenge of phase distortion.

Traditional speech enhancement approaches have primarily relied on statistical signal processing techniques. A seminal work in this field is the MMSE estimator developed by \citep{MMSE}, which performs clean speech spectrum estimation based on Gaussian distribution assumptions of both speech and noise signals. Although the MMSE algorithm demonstrates satisfactory performance in stationary noise environments, its effectiveness is constrained by the accuracy of a prior signal-to-noise ratio (SNR) estimation, often leading to musical noise artifacts and speech quality degradation under non-stationary noise conditions. As another classical approach, Wiener filtering \citep{wiener} achieves speech enhancement by constructing an optimal gain function based on power spectral density ratios, yet this method suffers from phase estimation bias and residual noise issues in dynamic acoustic environments. Furthermore, spectral subtraction \citep{spectralsubtraction}, as a computationally efficient enhancement technique, operates by directly subtracting noise spectrum estimates from noisy speech spectra, but this approach is prone to generate spectral distortion and artificial artifacts due to over-subtraction.

In recent years, the rapid development in deep learning have ushered in a paradigm shift for speech enhancement, with data-driven noise modeling approaches achieving remarkable performance improvements. Early investigations \citep{TCNN, TCN} predominantly employed convolutional neural network (CNN) architectures, which demonstrated substantial advancements over conventional signal processing methods by leveraging their exceptional capability in local spectral feature extraction. However, the inherent limitation of local receptive fields in CNNs posed significant challenges in modeling long-term temporal dependencies in speech signals, prompting the exploration of more expressive network architectures.

To address these limitations, researchers developed recurrent neural networks (RNNs) and their advanced variant, long short-term memory (LSTM) networks. These architectures incorporate gated mechanisms and sequential recursive connections to effectively model long-range temporal dependencies, significantly improving robustness in non-stationary noise environments \citep{Harmonic}. For instance, CRN \citep{CRN} integrates convolutional layers with recurrent structures to enable joint spatio-temporal feature learning. DCCRN \citep{DCCRN} further enhances spectral processing by introducing complex-valued operations, while DCCRN+ \citep{DCCRN+} combines complex TF-LSTM and subband processing to mitigate neural network-induced distortion. Despite these advancements, the sequential computation inherent in RNN-based approaches imposes constraints on computational efficiency and scalability, limiting their deployment in real-time applications.

The advent of the Transformer architecture \citep{transformer} provided a novel solution to these challenges by enabling parallel modeling of global contextual information through self-attention mechanisms. This innovation significantly improved computational efficiency while achieving remarkable enhancement performance \citep{DBAIAT, DPCFCS}. For instance, SE-LMA-Transformer \citep{SE-LMA-Transformer} leverages self-attention to effectively learn global structural features of speech signals. Conformer \citep{conformer} and SE-Conformer \citep{SE-Conformer} innovatively integrate CNNs with Transformers, enabling collaborative modeling of local features and global dependencies. However, these approaches still face significant challenges in jointly capturing the temporal dynamics and spectral structures of speech, which limits their enhancement performance and robustness.

Based on these findings, the two-stage Transformer architecture proposed by \citep{DPT-FSNet} partially alleviates incomplete information perception through a T-F alternating modeling strategy. Subsequent research such as DNSIP \citep{DNSIP} further enhanced system performance by optimizing intermediate enhancement layer structures. Notably, these performance improvements often come at the cost of dramatically increased computational complexity. Previous state-of-the-art (SOTA) models like MP-SENet \citep{MPSENET} typically require more than 64 input channels to achieve optimal performance, severely limiting their practical application in resource-constrained environments. Recent studies \citep{Mambaseunet} have also pointed out that conventional two-stage methods usually perform dimensionality reduction only along the frequency axis, and this single-scale feature processing approach struggles to adequately capture input characteristics at multiple resolutions, thereby constraining further model performance enhancement.

To overcome these challenges, our previous work \citep{MUSE} proposed the MUSE network, which innovatively combines sub-quadratic complexity Taylor multi-head self-attention (T-MSA) with a U-Net framework \citep{U-Net}, effectively alleviating the high computational overhead caused by the quadratic complexity relative to sequence length and multiple input channels in traditional self-attention mechanisms. By applying the Taylor-Transformer to capture multi-scale information at different resolutions, MUSE achieves competitive performance with only 16 input channels and low computational complexity. However, in-depth analysis reveals that the attention mechanism in MUSE primarily focuses on learning long-range, coarse-grained global dependencies at the spectral level, while its capability to model fine-grained local features in the time-frequency domain remains relatively limited, constraining the model's performance to some extent.

Building upon the above research challenges, as an extended and enhanced version of our prior conference work \citep{MUSE}, this study presents LORT, a lightweight single-channel speech enhancement model that also maintains outstanding performance utilizing only 16 input channels. Specifically, we design the Locally Refined Convolution (LRC) block, which integrates Convolutional Feedforward Networks (CFNs), Time-Frequency Dense Local Convolution (TF-DLC), and gated units, substantially improving fine-grained local feature modeling. Furthermore, we propose the Spatial-Channel Enhanced Attention (SCEA) mechanism, which jointly optimizes channel and spatial dimensions, thereby significantly enhancing the global representation capacity of the Taylor-Transformer and mitigating the limitations observed in MUSE.

The key contributions of this study can be summarized as follows:
\begin{itemize} 
\item We introduce a novel LRC block that synergistically integrates CFNs, TF-DLC, and gated units to effectively address the limitations of MUSE in fine-grained feature learning.
\item We develop a SCEA branch that enables coordinated optimization across channel and spatial dimensions, thereby enhancing the global modeling capability of the Taylor-Transformer. 
\item By incorporating LRC and SCEA into an enhanced Taylor-Transformer framework and integrating it with a U-Net-based multi-scale fusion structure, we construct an efficient and lightweight LORT model.
\item Comprehensive experiments on multiple benchmark datasets demonstrate that compared to previous SOTA methods, LORT achieves comparable or superior enhancement performance with significantly reduced parameters and computational costs.
\end{itemize}

The remainder of this paper is organized as follows: Section \ref{problem} describes the research problem in the field of speech enhancement. Section \ref{LORTMODEL} provides a comprehensive description of our proposed LORT model, including its core architecture, innovative modules, and technical implementation details. Section \ref{Experiments} presents the experimental setup, covering dataset composition, evaluation metrics, baseline methods, and implementation specifics. Section \ref{results} presents comprehensive results, including ablation studies and comparisons with previous SOTA methods. Finally, Section \ref{conclusions} concludes the research results of this paper, and discusses the future improvement and application prospects.

\section{Problem formulation}
\label{problem}

The monaural speech enhancement problem can be formally defined as the estimation of clean speech signal $s(t)$ from its noise-corrupted observation $y(t)$, where the noisy signal is typically modeled as the additive combination of clean speech and background noise:

\begin{equation}
y(t) = s(t) + n(t)
\end{equation}
where $n(t)$ represents the ambient noise component. In practical acoustic environments, $n(t)$ often exhibits non-stationary behavior with complex temporal-spectral characteristics.

Modern speech enhancement systems predominantly operate in the time-frequency (T-F) domain, which is typically obtained via the short-time Fourier transform (STFT). The STFT converts time-domain signals into time-frequency representations, which are critical for separating speech from noise.

The complex STFT representation of the noisy speech, $Y(m,k)$, can be further decomposed into magnitude $|Y(m,k)|$ and phase $\angle Y(m,k)$ components:
\begin{equation}
Y(m,k) = |Y(m,k)| e^{j\angle Y(m,k)}
\end{equation}
where the magnitude spectrum characterizes the energy distribution of speech in the T-F domain, while the phase spectrum plays a crucial role in preserving the naturalness and intelligibility of enhanced speech. However, due to the unstructured nature and high volatility of phase information, accurate phase estimation remains a formidable challenge in speech enhancement. Consequently, most T-F algorithms primarily focus on magnitude enhancement while retaining the noisy phase spectrum, which may fundamentally limit the performance ceiling. Recent advances in phase-aware speech enhancement, such as \citep{MPSENET}, have introduced dedicated phase decoders capable of explicitly enhancing phase spectra, thereby effectively alleviating this long-standing problem. The signal reconstruction in such frameworks is obtained via inverse STFT (iSTFT) as follows:

\begin{equation}
\hat{s}(t) = \text{iSTFT} \{|\hat{Y}(m,k)| \cdot e^{\angle \hat{S}(m,k)}\}
\end{equation}
where $|\hat{Y}(m,k)|$ denotes the enhanced magnitude spectrum, and $\angle \hat{S}(m,k)$ represents the estimated clean phase spectrum predicted by the model.

\begin{figure*}[t]
  \centering
  \includegraphics[width=0.99\linewidth]{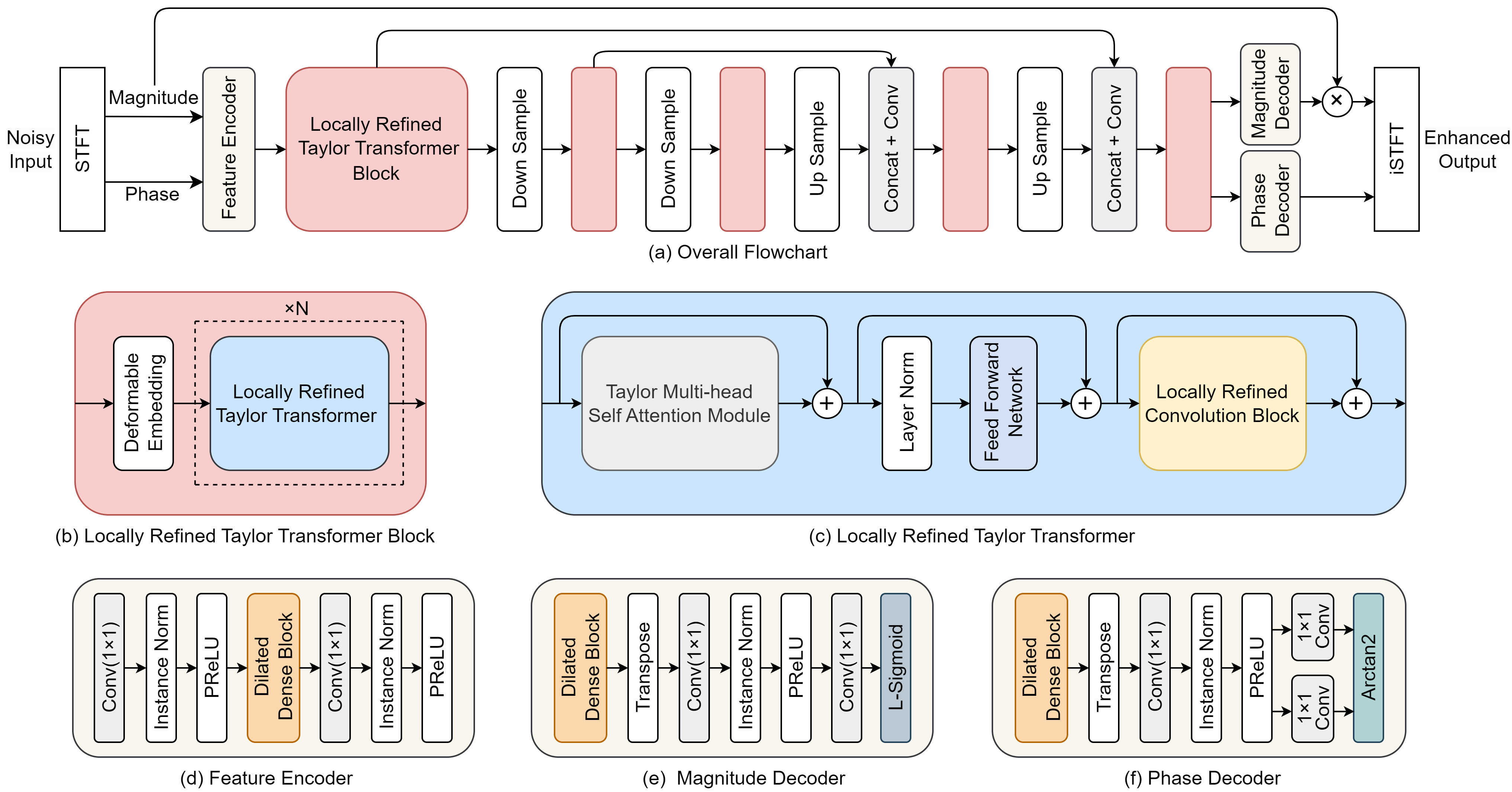}
  \caption{The overall architecture of proposed LORT model.}
  \label{figure1}
\end{figure*}

\section{Methodology}
\label{LORTMODEL}

\subsection{Model architecture}

The overall structure of LORT is illustrated in Figure \ref{figure1}. Given a noisy speech input \(x\), we apply STFT to obtain its magnitude spectrum \(X_m \in \mathbb{R}^{T \times F}\) and phase spectrum \(X_p \in \mathbb{R}^{T \times F}\), forming the input \(X \in \mathbb{R}^{T \times F \times 2}\) for the model. First, \(X\) is transformed into the corresponding features within an intermediate feature space through a feature encoder. Subsequently, these features are processed through multiple locally refined Taylor transformer blocks, with upsampling and downsampling operations applied between these blocks to alternately learn speech information at different resolutions. Finally, the enhanced magnitude spectrum \(X'_m \in \mathbb{R}^{T \times F}\) and phase spectrum \(X'_p \in \mathbb{R}^{T \times F}\) are independently decoded, and the denoised speech is obtained through inverse short-time Fourier transform (iSTFT).

\subsection{Encoder and decoder}

The encoder-decoder architecture of LORT is depicted in the lower part of Figure \ref{figure1}. Specifically, the feature encoder comprises two convolutional layers and a Dilated DenseNet \cite{DenseNet}. The initial convolutional layer expands the two input channels into an intermediate feature map with \(16\) channels, while the second convolutional layer reduces the frequency dimension by half to lower complexity. In this study, the Dilated DenseNet is configured with a depth of \(4\), and the dilation factor of each block is set to \(1, 2, 4, 8\), to effectively capture the spectral characteristics of the speech. Both the magnitude decoder and the phase decoder consist of a Dilated DenseNet identical to the encoder, followed by a 2-D transpose convolution and an output layer using 2-D convolution. The key difference between the two lies in the activation function. The magnitude decoder makes use of a learnable sigmoid function (L-Sigmoid) \citep{metricgan+}, while the phase decoder adopts an arctangent function (Arctan2) \cite{MPSENET}.

\subsection{Locally refined taylor transformer block}

The locally refined Taylor transformer block consists of a deformable embedding and \(N\) locally refined Taylor transformers. The deformable embedding, comes from \cite{taylor}, which is a combination of depthwise separable and deformable convolutions (DSDCN). Its purpose is to control the range of the receptive field through feature offset, enabling the capture of crescent-shaped spectral features as effectively as possible. The locally refined Taylor transformer comprises the Taylor multi-head self-attention module, layer normalization (LN), feedforward network, and LRC block, as shown in the middle part of Figure \ref{figure1}.

\subsubsection{Taylor multi-head self-attention module}

\begin{figure}[t]
  \centering
  \includegraphics[width=0.99\linewidth]{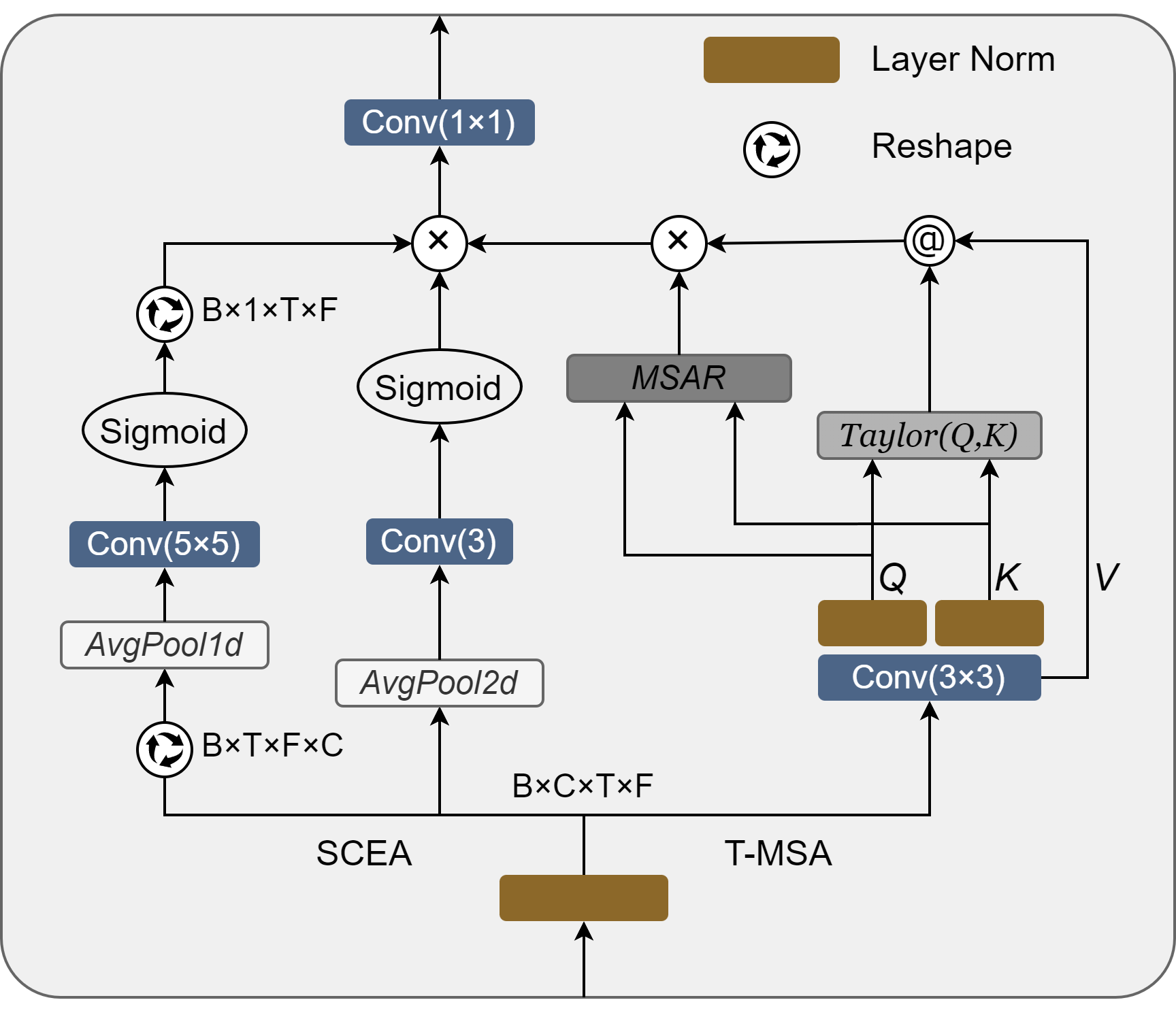}
  \caption{Taylor Multi-head Self-attention Module. The symbol $@$ denotes matrix multiplication.}
  \label{figure2}
\end{figure}

Inspired by \citep{MUSE}, the Taylor multi-head self-attention module is composed of two branches, as illustrated in Figure \ref{figure2}. The first branch, T-MSA, emphasizes global attention while reducing cross-channel attention. The second branch, SCEA, compensates for the lack of channel information exchange in T-MSA and enhances the sensitivity of the model to significant spatial information.

\textbf{T-MSA.} For general multi-head self-attention (MHSA), we have the following formula:
\begin{align}
  V' = \text{Softmax}\left(\frac{Q^{T}K}{\sqrt{D}}\right)V^T
\end{align}
Given that the softmax function can be represented as:
\begin{align}
  \text{Softmax}(z_i) = \frac{e^{z_i}}{\sum_{j=1}^n e^{z_j}} \in (0,1)
\end{align}
We can thus rewrite equation (1) in an alternative form:
\begin{align}
  V'_i = \frac{\sum_{j=1}^N \text{exp}\left(\frac{Q_{i}^{T}K_j}{\sqrt{D}}\right)V_{j}^{T}}{\sum_{j=1}^N \text{exp}\left(\frac{Q_{i}^{T}K_j}{\sqrt{D}}\right)}
\end{align}
where the matrix indexed by \(i\) denotes the vector in the \(i\)-th row of the matrix. By applying the first-order Taylor expansion of \text{exp} at 0, $Taylor(Q_i,K_j) = 1+\left(\frac{Q_{i}^{T}K_j}{\sqrt{D}}\right)+o\left(\frac{Q_{i}^{T}K_j}{\sqrt{D}}\right) \approx \text{exp}\left(\frac{Q_{i}^{T}K_j}{\sqrt{D}}\right)$, we can rewrite equation (3) as:
\begin{align}
  V'_i = \frac{\sum_{j=1}^N(1+Q_{i}^{T}K_j+o(Q_{i}^{T}K_j))V_{j}^{T}}{\sum_{j=1}^N(1+Q_{i}^{T}K_j+o(Q_{i}^{T}K_j))}
\end{align}

For an input of \( t \times f \) patches, the computational complexity of the standard MHSA module and the T-MSA module is expressed as follows:
\begin{align}
  O(\textnormal{MHSA}) = 4tfD^2 + 2t^2f^2D \\
  O(\textnormal{T-MSA})=18tfD + 2tfD^2
\end{align}
where \(D\) represents the hidden layer dimension. Generally, \(D\) is much smaller than  \( t \times f \). Consequently, T-MSA requires fewer computational resources compared to standard MHSA, and this reduction in computational complexity becomes more obvious as \( t \times f \) increases.

Since the use of the Taylor expansion in T-MSA neglects the Peano remainder term, it introduces a certain degree of approximation error. To address this issue, we employ the Multi-scale Attention Refinement (MSAR) method, as described in \cite{taylor}, which learns the local information of the \(Q\) and \(K\) matrices to rectify the inaccurate output \(V'\).

\textbf{SCEA.} Inspired by \cite{ARFDCN}, we design the SCEA branch to guide the network in capturing inter-channel feature relationships and to focus on significant regions of the phase and magnitude spectra. For the channel branch, we initially employ 2-D pooling to aggregate the T-F features of speech, followed by a 1-D convolution with a kernel size of \(3\) to facilitate inter-channel information exchange. For the spatial branch, we first apply 1-D pooling to aggregate the channel information of speech, followed by a $5 \times 5$ convolution to capture spatial information at different scales and generate the corresponding spatial feature maps.

\begin{figure}[t]
  \centering
  \includegraphics[width=0.99\linewidth]{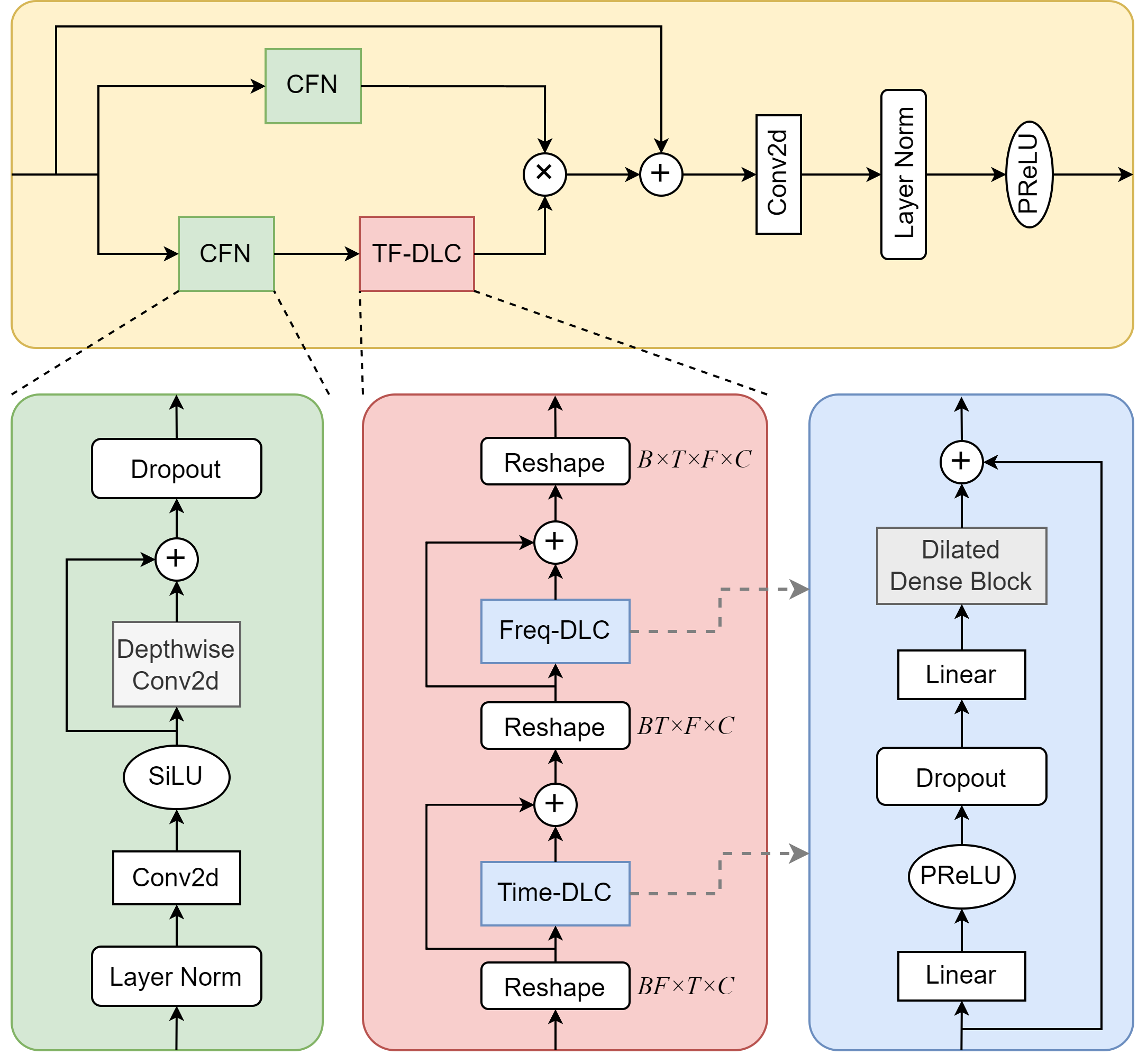}
  \caption{The flowchart of the presented Locally Refined Convolution Block (LRC).}
  \label{figure3}
\end{figure}

\subsubsection{Locally refined convolution block}

The core component of the LRC block is a gated unit composed of CFN and TF-DLC. CFN primarily assists the gated unit in capturing spatially relevant local patterns, while TF-DLC focuses on modeling fine-grained local information. To accelerate model convergence, a residual connection \cite{resnet} is incorporated between the input and output of the gated unit. The overall structure is illustrated in Figure \ref{figure3}.

\textbf{CFN.} Instead of the traditional gated unit, which often employs a combination of linear and activation functions, this block first processes the sequence through a LN layer, followed by a $1 \times 1$ convolution and SiLU activation \cite{SiLU}. Finally, through a 2-D depthwise convolution with a residual connection, it preliminarily models local features. 

\textbf{TF-DLC.} We sequentially employ two DLC blocks to model fine-grained local information in the time and frequency dimensions. Each DLC block begins with two linear layers, similar to a conventional feedforward network \cite{transformer}. This is followed by a dilated dense block with dense connections, designed to improve information flow within the network, expand the receptive field, and promote information aggregation at different resolutions. To facilitate gradient propagation, a residual connection is also introduced between the initial input and the model output.

\subsection{Loss function}

Effective loss function design is crucial for optimizing the performance of speech enhancement models. In our previous study \citep{Primek}, we observed that directly applying waveform-based time loss in the time-frequency (T-F) domain often leads to suboptimal performance. To address this, we formulate a multi-component loss function that refines the enhanced spectrogram by incorporating complex loss $\mathcal{L}_{\text{RI}}$, magnitude loss $\mathcal{L}_{\text{Mag.}}$, phase loss $\mathcal{L}_{\text{Pha.}}$ and STFT consistency loss $\mathcal{L}_{\text{Con.}}$. These components are described as shown below:

The complex loss and magnitude loss are designed to minimize the $L2$ distance between the predicted and ground-truth spectrograms in the complex domain:
\begin{equation}
\begin{aligned}
&\mathcal{L}_{\text{RI}} = \mathbb{E}_{Y_r, \hat{Y}_r}[\lVert Y_r - \hat{Y}_r \rVert^{2}] + \mathbb{E}_{Y_i, \hat{Y}_i}[\lVert Y_i - \hat{Y}_i \rVert^{2}] \\
&\mathcal{L}_{\text{Mag.}} = \mathbb{E}_{Y_m, \hat{Y}_m}[\lVert Y_m - \hat{Y}_m \rVert^{2}] \\
\end{aligned}
\end{equation}

Since accurate phase estimation remains a challenging problem in speech enhancement, we incorporate an inverse-wrapping phase loss proposed in \citep{MPSENET}:
\begin{equation}
\begin{aligned}
&\mathcal{L}_{\text{Pha.}} = \mathcal{L}_{\text{GD}} + \mathcal{L}_{\text{IAF}} + \mathcal{L}_{\text{IP}}
\end{aligned}
\end{equation}
where \text{GD}, \text{IAF} and \text{IP} denote group delay, instantaneous angular frequency and instantaneous phase respectively. Furthermore, as STFT and its inverse transformation (iSTFT) introduce reconstruction bias, we employ a consistency loss \citep{SCPGAN} to ensure that the deviation introduced when applying the STFT is minimized:
\begin{equation}
\begin{aligned}
\mathcal{L}_{\text{Con.}} &= \mathbb{E}_{\hat{Y}_r}[\lVert \hat{Y}_r - \text{STFT}(\text{iSTFT}(\hat{Y}_r))\rVert^{2}] + \\
 &\mathbb{E}_{\hat{Y}_i}[\lVert \hat{Y}_i - \text{STFT}(\text{iSTFT}(\hat{Y}_i))\rVert^{2}] \\
\end{aligned}
\end{equation}

While the aforementioned losses ensure spectral fidelity, optimizing perceptual quality metrics such as PESQ and STOI remains challenging due to their non-differentiability. To bridge this gap, we introduce a metric-guided adversarial loss, where a learnable discriminator serves as a surrogate quality assessment module. This enables the generator to align its optimization process more closely with human auditory perception. Specifically, the generator loss $\mathcal{L}{\text{G}}$ encourages the network to produce high-quality enhanced speech, while the discriminator loss $\mathcal{L}{\text{D}}$ refines the quality assessment by incorporating a PESQ-based constraint:
\begin{equation}
\begin{aligned}
\mathcal{L}&_{\text{G}} = \mathbb{E}_{Y_m, \hat{Y}_m}[\lVert D(Y_m, \hat{Y}_m) - 1 \rVert^{2}] \\
\mathcal{L}&_{\text{D}} = \mathbb{E}_{Y_m}[\lVert D(Y_m, Y_m) - 1 \rVert^{2}] \\ 
& + \mathbb{E}_{Y_m, \hat{Y}_m}[\lVert D(Y_m, \hat{Y}_m) - Q_{\text{PESQ}} \rVert^{2}] 
\end{aligned}
\end{equation}
where \(G\) represents the generator, \(D\) denotes the discriminator, and the range of PESQ score is normalized to $Q_{\text{PESQ}} \in [0,1]$. 

The final loss function $\mathcal{L}_{\text{Total}}$ is formulated as follows:
\begin{align}
  \mathcal{L}_{\text{Total}} = \alpha_{1}\mathcal{L}_{\text{RI}} + \alpha_{2}\mathcal{L}_{\text{Mag.}} + \alpha_{3}\mathcal{L}_{\text{Pha.}} + \alpha_{4}\mathcal{L}_{\text{Con.}} + \alpha_{5}\mathcal{L}_{\text{G}}
\end{align}
where $\alpha_{1}$, $\alpha_{2}$, $\alpha_{3}$, $\alpha_{4}$, $\alpha_{5}$ are hyperparameters. This hierarchical loss structure ensures both spectral accuracy (lower-level objective) and perceptual quality (higher-level objective) are optimized incrementally.

\section{Experiments}
\label{Experiments}

\subsection{Dataset}

To comprehensively evaluate the effectiveness of our proposed model across diverse acoustic conditions, we conduct experiments on two benchmark datasets: VCTK+DEMAND \citep{vctk+demand} and DNS Challenge 2020.

VCTK+DEMAND Dataset: This mixed dataset is designed to facilitate comprehensive speech enhancement research. The clean speech data is sourced from the VoiceBank corpus \cite{voicebank}, which is characterized by high-fidelity recordings from a diverse group of speakers. The mixed noise is derived from the DEMAND dataset \cite{Demand}, a comprehensive collection of real-world noise recordings. The generated mixed speech dataset consists of 12,396 utterances contributed by 30 speakers. To ensure effective model training and unbiased evaluation, the dataset is partitioned based on speakers. Specifically, data from 28 speakers are utilized for the training set, while the remaining 2 speakers are reserved for testing set. In terms of noise characteristics, the training set includes 10 types of noise with signal-to-noise ratios (SNRs) ranging from 0 to 15 dB, and the testing set includes 5 types of noise with SNRs spanning from 2.5 to 17.5 dB.

DNS Challenge 2020 Dataset: As one of the most comprehensive public speech enhancement benchmarks for speech enhancement research, this dataset provides extensive coverage of real-world noise conditions, featuring approximately 2150 speakers and 150 types of noise, which ensures exceptional diversity \citep{dns2020}. 
For our experiments, we constructed the dataset through random sampling. The training set comprises 50000 speech data samples with a uniform distribution of SNR ranging from -5 dB to 15 dB (2500 samples per 1 dB interval). This uniform distribution of samples across different SNR values ensures that the model receives balanced training across a wide spectrum of noise levels.
The validation set follow the same SNR range as the training set but contains 5000 speech samples (250 samples per 1 dB interval).
The testing set includes a total of 5000 speech samples. The SNRs of the testing set are specifically set at -5 dB, 0 dB, 5 dB, 10 dB, and 15 dB, with 1000 speech data samples for each SNR level. These selected SNR values represent common real-world noisy scenarios, allowing us to rigorously evaluate the model's performance under well-defined and practical conditions.

To maintain consistency and compatibility with common speech processing systems, all samples in the two datasets are downsampled to 16 kHz.

\subsection{Baselines}

To rigorously evaluate the performance of the proposed speech enhancement model, we compare it against six advanced baseline systems.

TSTNN \citep{TSTNN} is a time-domain speech enhancement model employing a two-stage architecture. Unlike T-F domain methods, it processes raw waveform inputs directly without applying STFT. The model features a dual-branch design: one branch focuses on capturing local temporal structures, while the other encodes long-range contextual information. This design enables effective modeling of both short-term dynamics and global dependencies in speech signals.

DCCRN \citep{DCCRN} pioneers the use of complex-valued operations by integrating complex convolutional layers and complex LSTM networks. While it performs inference in the complex domain, the model ultimately predicts only the magnitude spectrum, leaving phase components untouched.

FRCRN \citep{FRCRN} builds upon DCCRN by incorporating a Frequency Recurrent Feedforward Sequential Memory Network (FSMN) layer. This enhancement enables the convolutional-recurrent block to model both local time-frequency structures and long-range frequency dependencies. However, the model remains constrained to magnitude-based spectral estimation in the complex domain.

CMGAN \citep{CMGAN} is a GAN-based architecture that alternately models time-domain and frequency-domain features through a two-stage Conformer encoder. The decoder branches separately estimate magnitude and complex spectra to ensure more stable and accurate waveform reconstruction. The discriminator, similar to MetricGAN, leverages PESQ scores as supervision and integrates perceptual consistency loss to further enhance perceptual quality.

MP-SENet \citep{MPSENET} extends CMGAN by replacing the complex decoder with a dedicated phase decoder, allowing the model to explicitly predict enhanced phase information rather than relying on implicit modeling. This architectural refinement significantly improves the perceptual realism of the reconstructed signal.

MUSE \citep{MUSE}, a lightweight speech enhancement model we previously proposed, differs from the popular two-stage designs. Built upon a U-Net backbone enhanced with a Multipath Enhanced Taylor (MET) Transformer module, MUSE can simultaneously model spectral features across frequency bands and spatial representations. This allows for a more comprehensive analysis and enhancement of speech signals in both frequency and spatial dimensions.

\subsection{Experimental setup}

The FFT length is set to 510, with a window length of 510 and a hop size of 100. The number of locally refined Taylor transformer is 4 (\(N = 4\)). In the DLC, the dilated dense block is configured with a depth of 2, a dilation factor of 2, and a convolutional kernel size of (19,1), with each convolution followed by InstanceNorm2d 
 (\cite{Instancenorm}) and PReLU activation (\cite{prelu}). The model undergoes 100 epochs of training, using the AdamW optimizer (\cite{adamw}), with an initial learning rate of 5e-4 and a decay factor of 0.99. The hyperparameters $\alpha_{1}$, $\alpha_{2}$, $\alpha_{3}$, $\alpha_{4}$, $\alpha_{5}$ are set to 0.1, 0.9, 0.3, 0.1, and 0.05, respectively.

\subsection{Evaluation metrics}

To comprehensively assess the performance of speech enhancement models, we employ three widely adopted objective and subjective metrics. (1) Perceptual Evaluation of Speech Quality (PESQ) \citep{pesq} is designed to evaluate the perceptual quality of enhanced speech by comparing it to the clean reference signal, with a score range from -0.5 to 4.5; (2) Short-Time Objective Intelligibility (STOI) \citep{stoi} measures the proportion of intelligible content in speech signals by analyzing time-frequency correlations, with a score range from 0 to 1; (3) Mean Opinion Score (MOS) metrics provide subjective yet highly informative assessments of speech quality from a human listener's perspective. These include signal distortion prediction (CSIG), background noise interference prediction (CBAK), and overall speech quality prediction (COVL) \citep{csig}, all rated from 1 to 5. In addition to these performance metrics, we add FLOPs, a measure of computational complexity, based on the processing of a 2-second, 16kHz speech sample on the GPU.

\section{Results and analysis}
\label{results}

We begin our evaluation by conducting a comprehensive ablation study to assess the effectiveness of each component in the proposed method. Subsequently, we compare our model with a set of well-established baselines on the widely adopted VCTK+DEMAND dataset to evaluate general speech enhancement performance under diverse real-world noise conditions. Finally, to further validate the robustness and scalability of our approach, we perform extensive experiments on the DNS Challenge 2020 dataset, which contains a larger volume of data and a broader range of noise types.

\subsection{Ablation studies}

To systematically evaluate the contribution of each architectural component in our proposed LORT framework, we conduct ablation studies on the VCTK+DEMAND dataset by selectively removing or replacing individual modules, as summarized in Table~\ref{tab1}. In this table, “w/o” refers to the removal of a module, while “A→B” indicates the replacement of component B with A.

The complete LORT model achieves the best overall performance across all metrics, while maintaining a compact model size of only 0.96M parameters. Removal of the SCEA module leads to noticeable performance drops, particularly in PESQ (0.06 drop) and COVL (0.07 drop), highlighting the importance of SCEA as a complementary branch to the T-MSA module. This result suggests that SCEA significantly enhances the model’s capacity to model global dependencies across time and frequency domains.

\begin{table}[t]
  \caption{Comparison of ablation study results for LORT and its variants.}
  \label{tab1}
  \centering
  \tabcolsep=0.08cm
  \begin{tabular*}{\hsize}{@{\hspace{0.1cm}}@{\extracolsep{\fill}}lcccccc@{}}
    \hline
    \textbf{Method} & \textbf{PESQ} & \textbf{CSIG} & \textbf{CBAK} & \textbf{COVL} & \textbf{\# Params} \\
    \hline
    LORT  & \textbf{3.51} & \textbf{4.74} & \textbf{3.91} & \textbf{4.23} & 0.96M  \\
    w/o SCEA  & 3.44  & 4.70 & 3.86 & 4.16  & 0.96M                    \\
    w/o CFN & 3.45 & 4.69 & 3.84 & 4.17  & 0.75M     \\
    Linear → CFN   & 3.46  & 4.70 & 3.86 & 4.19  & 0.79M                   \\
    w/o TF-DLC & 3.40 & 4.67 & 3.80 & 4.12 & 0.70M      \\
    Conv. → LRC   & 3.46 & 4.70 & 3.88 & 4.19 & 0.96M                    \\
    \hline
  \end{tabular*}
\end{table}

\begin{table}[t]
  \caption{Test results for different configurations. N=4, C=16 are the default LORT configurations for our other experiments.}
  \label{tab2}
  \centering
  \tabcolsep=0.2pt
  \tabcolsep=0.02cm
  \begin{tabular*}{\hsize}{@{\hspace{0.1cm}}@{\extracolsep{\fill}}lccccccc@{}}
    \hline
    \textbf{N} & \textbf{C} & \textbf{PESQ} & \textbf{CSIG} & \textbf{CBAK} & \textbf{COVL} & \textbf{\# Params} & \textbf{FLOPs} \\
    \hline
    1  & 16  & 3.38  & 4.66 & 3.80 & 4.10  & 0.56M  & 13.44G                  \\
    2  & 16  & 3.45  & 4.71 & 3.87 & 4.18  & 0.69M  & 14.57G                  \\
    3  & 16  & 3.49  & 4.73 & 3.89 & 4.21  & 0.82M  & 15.70G                  \\
    4  & 16  & 3.51  & 4.74 & 3.91 & 4.23  & 0.96M  & 16.83G                  \\
    5  & 16  & 3.51  & 4.75 & 3.90 & 4.22  & 1.09M  & 17.96G                  \\
    4  & 8  & 3.44  & 4.68 & 3.84 & 4.17  & 0.33M  & 5.82G                  \\
    4  & 24  & \textbf{3.53} & \textbf{4.76} & \textbf{3.93} & \textbf{4.25} & 1.89M & 33.04G  \\
    \hline
  \end{tabular*}
\end{table}

\begin{table*}[ht]
\begin{center}
  \caption{Performance comparison on VCTK+DEMAND dataset. "-" denotes the data is not provided in the original paper.}
  \setlength{\tabcolsep}{4mm}
  \label{tab3}
  \centering
  \tabcolsep=3pt
  \begin{tabular}{lcccccccc}
    \hline
    \textbf{Model}  & \textbf{Year}  & \textbf{\# Params}  & \textbf{FLOPs}  & \textbf{PESQ}  & \textbf{STOI}  &\textbf{CSIG}   & \textbf{CBAK}   & \textbf{COVL}  \\
    \hline
    Noisy  & - & - & - & 1.97 & 0.91 & 3.35 & 2.44 & 2.63                   \\
    \hline
    Wiener \citep{wiener} &1998 & - & - &2.22 & 0.91 & 3.23  & 2.68 & 2.67 \\
    Wave-U-Net \citep{waveunet} &2018  & 10.4M & - &2.4 & 0.93 &3.52  & 3.24 & 2.96  \\
    MetricGAN \citep{MetricGAN} & 2019 & - & - & 2.86 & - & 3.99 & 3.18 & 3.42                           \\
    PHASEN \citep{phasen} & 2020 & 8.76M & - & 2.99 & - & 4.21 & 3.55 & 3.62                                 \\
    TSTNN \citep{TSTNN} & 2021 & 0.92M & - & 2.96 & 0.95 & 4.33 & 3.53 & 3.67  \\
    MetricGAN+ \citep{metricgan+} & 2021 & - & - & 3.15 & - & 4.14 & 3.16 & 3.64                      \\
    SE-Conformer \citep{SE-Conformer} & 2021 & - & - & 3.13 & 0.95 & 4.45 & 3.55 & 3.82       \\
    DPT-FSNet \citep{DPT-FSNet} & 2021  & 0.91M & - & 3.33 & \textbf{0.96} & 4.58 & 3.72 & 4.00    \\
    DB-AIAT \citep{DBAIAT} & 2022 & 2.81M & 110.20G & 3.31 & \textbf{0.96} & 4.61 & 3.75 & 3.96     \\
    CMGAN \citep{CMGAN} & 2022  & 1.83M & 63.15G & 3.41 & \textbf{0.96} & 4.63 & 3.94 & 4.12         \\
    TridentSE \citep{Trident}  & 2023 & 3.03M & - & 3.47 & 0.96 & 4.70 & 3.81 & 4.10            \\
    DPCFCS-Net \citep{DPCFCS} & 2023 & 2.86M & 130.65G & 3.42 & \textbf{0.96} & 4.71 & 3.88 & 4.15  \\
    MP-SENet \citep{MPSENET} & 2023 & 2.05M & 74.29G & 3.50 & \textbf{0.96} & 4.73 & 3.95 & 4.22          \\
    stDPT \citep{stDPT} & 2023 & 1.14M & - & 3.13 & - & 4.50 & 3.78 & 3.89            \\
    S4ND U-Net \citep{S4NDU-Net} & 2023 & 0.75M & - & 3.15 & - & 4.52 & 3.62 & 3.85            \\
    SE-LMA-Transformer \citep{SE-LMA-Transformer} & 2024 & 5.60M & - & 3.40 & 0.96 & 4.65 & 3.87 & 4.12      \\
    MUSE \citep{MUSE} & 2024 & 0.51M & 9.43G & 3.37 & 0.95 & 4.63 & 3.80 & 4.10            \\
    SEMamba \citep{semamba} & 2024 & 2.25M & 65.46G & \textbf{3.52} & \textbf{0.96} & \textbf{4.75} & \textbf{3.98} & \textbf{4.26}            \\
    SINAI-MoSE\citep{SINAI-MoSE}  &2025  & 1.52M & - & 3.26 & 0.96  & 4.59 &3.83  &4.03 \\
    \hline
    \textbf{LORT} & 2025 & 0.96M & 16.83G & 3.51 & \textbf{0.96} & 4.74 & 3.91 & 4.23                   \\
    \hline
  \end{tabular}
\end{center}
\end{table*}

Eliminating the CFN component results in a decrease in all evaluation indicators, with PESQ decreasing from 3.51 to 3.45, which indicates that CFN plays a crucial role in the TF-DLC module. Replacement with a simple linear projection further confirms CFN's superiority in modeling inter-frequency correlations (PESQ drops to 3.46). Complete removal of the TF-DLC block causes the most severe performance decline (COVL decreases by 0.11 points), underscoring its fundamental role in joint time-frequency context encoding. Furthermore, replacing the LRC block with the standard convolutional module from Conformer also leads to noticeable decreases, validating the superiority of the proposed LRC in capturing fine-grained local structures (PESQ reduction of 0.05).

Next, we further analyze the architecture of the model by varying the number of locally refined taylor transformer blocks ($N$) and the output channel dimension ($C$) of the encoder. The evaluation results are presented in Table~\ref{tab2}, with comparisons in terms of four standard metrics (PESQ, CSIG, CBAK, and COVL), model size, and FLOPs.

We first fix the encoder channel dimension ($C = 16$) and vary the number of LORT blocks ($N$) from 1 to 5. The performance consistently improves as $N$ increases from 1 to 4, indicating that deeper stacking of LORT blocks enhances the model’s capacity to capture multi-scale contextual features. Specifically, the PESQ score increases from 3.38 to 3.51, while COVL rises from 4.10 to 4.23. However, with a further increase in $N = 5$, although the number of parameters (1.09M) and computational cost (17.96G FLOPs) increase, only CSIG shows a slight improvement, while other metrics such as CBAK and COVL exhibit subtle degradation. This suggests that overly deep configurations may lead to overfitting or optimization difficulties, without further performance gain.

We then fix $N = 4$ and adjust the encoder output dimension ($C$) to assess the effect of feature representation capacity. Reducing $C$ to 8 substantially lowers both parameter count (0.33M) and FLOPs (5.82G), yet leads to noticeable performance degradation such as PESQ drops to 3.44. In contrast, increasing $C$ to 24 results in the highest overall performance (PESQ 3.53, CSIG 4.76, CBAK 3.93, COVL 4.25), albeit with a significant rise in model size (1.89M) and computation (33.04G). These findings reveal a clear trade-off between model complexity and enhancement quality.

\begin{table}[t]
  \caption{Performance comparison of LORT with different STFT hop sizes.}
  \label{tabhopsize}
  \centering
  \setlength{\tabcolsep}{4pt} 
  \begin{tabular}{lccccc} 
    \hline
    \textbf{Hop Size} & \textbf{PESQ} & \textbf{CSIG} & \textbf{CBAK} & \textbf{COVL} & \textbf{FLOPs} \\
    \hline
    100  & \textbf{3.51} & \textbf{4.74} & \textbf{3.91} & \textbf{4.23} & 16.83G  \\
    120  & 3.49  & 4.72 & 3.88 & 4.21  & 13.55G                  \\
    150  & 3.46  & 4.71 & 3.86 & 4.18 & 11.15G                  \\
    \hline
  \end{tabular}
\end{table}

Finally, we investigate the impact of the STFT hop size on model performance, as this parameter directly influences the trade-off between temporal resolution and computational efficiency. The evaluation results for three different hop sizes are presented in Table~\ref{tabhopsize}.

Our findings indicate that a hop size of 100 achieves the best performance across all metrics. This can be attributed to its finer temporal resolution, which is crucial for preserving the transient speech components that are essential for intelligibility and naturalness. Increasing the hop size to 120 reduces the computational complexity (FLOPs drop from 16.83G to 13.55G) but leads to a performance degradation (PESQ decreases by 0.02). Further increasing the hop size to 150 results in a more significant drop in performance (PESQ=3.46) and even lower FLOPs (11.15G). This demonstrates that a coarser temporal resolution compromises the model's ability to capture fine-grained time-frequency dynamics. These results confirm that the choice of hop size involves a balance between efficiency and temporal precision. Considering the performance gains, a hop size of 100 is the optimal choice for the LORT model.

\subsection{Comparison with public methods on VCTK+DEMAND dataset}

Table~\ref{tab3} presents a comprehensive comparison between our proposed LORT model and a wide range of baseline systems, encompassing both classical methods like PHASEN, TSTNN, and MetricGAN+, and recent state-of-the-art (SOTA) models such as MP-SENet, DPCFCS-Net and SE-LMA-Transformer. Several light-weight architectures, namely MUSE and S4ND U-Net, are also included to highlight the efficiency-performance trade-off in low-complexity scenarios.

\begin{table}[t]
  \caption{Comparative Performance Analysis of LORT, MP-SENet (S), and MUSE (L).}
  \label{tab4}
  \centering
  \tabcolsep=0.2pt
  \tabcolsep=0.02cm
  \begin{tabular*}{\hsize}{@{\hspace{0.1cm}}@{\extracolsep{\fill}}lccccccc@{}}
    \hline
    \textbf{Method} & \textbf{PESQ} & \textbf{CSIG} & \textbf{CBAK} & \textbf{COVL} & \textbf{\# Params} & \textbf{FLOPs} \\
    \hline
    LORT  & \textbf{3.51} & \textbf{4.74} & \textbf{3.91} & \textbf{4.23} & 0.96M & 16.83G  \\
    MP-SENet (S)   & 3.42  & 4.68 & 3.86 & 4.16  & 1.16M & 42.11G                   \\
    MUSE (L)  & 3.40  & 4.66 & 3.84 & 4.13  & 1.14M  & 17.50G                  \\
    \hline
  \end{tabular*}
\end{table}

\begin{table*}[htb]
  \centering
  \caption{Performance comparison with baselines against DNS noise, representative BABBLE, and FACTORY noise on DNS Challenge dataset.}
  \begin{tabular}{c|c|c|ccccc|c|ccccc|cc}
    \hline
     \multirow{2}{*}{} & \multirow{2}{*}{\textbf{Model}} & \multirow{2}{*}{\textbf{\# Params}} & \multicolumn{5}{c|}{\textbf{PESQ}} & \multirow{2}{*}{\textbf{AVG}} & \multicolumn{5}{c|}{\textbf{STOI (\%)}} & \multirow{2}{*}{\textbf{AVG}} \\ 
    & & & -5dB & 0dB & 5dB & 10dB & 15dB & & -5dB & 0dB & 5dB & 10dB & 15dB &  \\
    \hline
    \multirow{10}{*}{\rotatebox[origin=c]{90}{\textbf{DNS}}} 
    & Noisy & - & 1.11 & 1.14 & 1.24 & 1.45 & 1.86 & 1.36 & 68.4 & 77.6 & 84.9 & 90.9 & 95.8 & 83.5  \\
    & TSTNN & 0.92M & 1.67 & 2.19 & 2.63 & 3.05 & 3.48 & 2.60 & 82.1 & 89.2 & 93.6 & 96.1 & 98.1 & 91.8  \\
    & DCCRN & 3.67M & 1.46 & 1.99 & 2.42 & 2.84 & 3.28 & 2.40 & 81.3 & 88.5 & 93.5 & 95.9 & 97.9 & 91.4  \\
    & FRCRN & 6.90M & 1.80 & 2.23 & 2.67 & 3.09 & 3.52 & 2.66  & 83.3 & 90.2 & 94.3 & 96.4 & 98.2 & 92.5  \\
    & CMGAN & 1.92M & 1.97 & 2.49 & 2.90 & 3.33 & 3.75 & 2.89 & 87.9 & 92.7 & 95.5 & 97.3 & 98.5 & 94.4  \\
    & MP-SENet & 2.05M & 2.15 & 2.65 & 3.05 & 3.47 & 3.88 & 3.04 & \textbf{89.0} & \textbf{93.4} & 96.0 & \textbf{97.6} & \textbf{98.6} & \textbf{95.0}  \\
    & MUSE & 0.51M & 1.75 & 2.20 & 2.61 & 3.06 & 3.50 & 2.62 & 82.9 & 89.8 & 93.9 & 96.3 & 98.1 & 92.2  \\
    & MP-SENet (S) & 1.16M & 2.01 & 2.52 & 2.93 & 3.35 & 3.77 & 2.92 & 88.1 & 92.8 & 95.6 & 97.3 & 98.5 & 94.5  \\
    & MUSE (L) & 1.14M & 1.81 & 2.24 & 2.69 & 3.14 & 3.58 & 2.69 & 83.4 & 90.3 & 94.3 & 96.4 & 98.2 & 92.5   \\
    \hline
    & \textbf{LORT} & 0.96M & \textbf{2.26} & \textbf{2.75} & \textbf{3.14} & \textbf{3.53} & \textbf{3.92} & \textbf{3.12} & \textbf{89.0} & \textbf{93.4} & \textbf{96.1} & \textbf{97.6} & \textbf{98.6} & \textbf{95.0}  \\
    \hline
    \multirow{10}{*}{\rotatebox[origin=c]{90}{\textbf{Babble}}} 
    & Noisy & - & 1.08 & 1.09 & 1.17 & 1.34 & 1.69 & 1.27 & 55.2 & 68.1 & 79.8 & 88.6 & 94.2 & 77.2  \\
    & TSTNN & 0.92M & 1.14 & 1.43 & 2.08 & 2.53 & 2.90 & 2.02 & 57.7 & 79.8 & 87.7 & 93.2 & 95.8 & 82.8  \\
    & DCCRN & 3.67M & 1.11 & 1.34 & 1.95 & 2.38 & 2.74 & 1.90 & 56.6 & 77.1 & 85.5 & 91.8 & 95.2 & 81.2  \\
    & FRCRN & 6.90M & 1.16 & 1.49 & 2.14 & 2.62 & 3.03 & 2.09 & 62.1 & 81.9 & 90.4 & 94.2 & 96.5 & 85.0  \\
    & CMGAN & 1.92M & 1.20 & 1.56 & 2.27 & 2.81 & 3.26 & 2.22 & 70.2 & 85.1 & 92.3 & 95.5 & 97.2 & 88.1  \\
    & MP-SENet & 2.05M & 1.27 & 1.83 & 2.50 & 2.99 & 3.42 & 2.40 & \textbf{72.5} & \textbf{86.7} & \textbf{93.4} & 96.1 & 97.8 & \textbf{89.3}  \\
    & MUSE & 0.51M & 1.15 & 1.46 & 2.10 & 2.57 & 2.95 & 2.05 & 58.6 & 81.2 & 89.8 & 93.8 & 96.3 & 83.9  \\
    & MP-SENet (S) & 1.16M & 1.23 & 1.68 & 2.30 & 2.84 & 3.30 & 2.27 & 70.6 & 85.4 & 92.5 & 95.6 & 97.3 & 88.3  \\
    & MUSE (L) & 1.14M & 1.19 & 1.53 & 2.18 & 2.68 & 3.11 & 2.14 & 64.2 & 83.4 & 90.9 & 94.5 & 96.7 & 85.9  \\
    \hline
    & \textbf{LORT} & 0.96M & \textbf{1.37} & \textbf{1.97} & \textbf{2.62} & \textbf{3.11} & \textbf{3.51} & \textbf{2.52} & \textbf{72.5} & \textbf{86.7} & \textbf{93.4} & \textbf{96.2} & \textbf{97.9} & \textbf{89.3}  \\
    \hline
    \multirow{10}{*}{\rotatebox[origin=c]{90}{\textbf{Factory}}} 
    & Noisy & - & 1.05 & 1.07 & 1.10 & 1.22 & 1.49 & 1.19 & 54.6 & 67.6 & 79.3 & 88.6 & 93.6 & 76.7  \\
    & TSTNN & 0.92M & 1.21 & 1.59 & 2.17 & 2.62 & 2.96 & 2.11 & 64.9 & 81.7 & 89.2 & 93.0 & 96.1 & 85.0  \\
    & DCCRN & 3.67M & 1.16 & 1.51 & 2.07 & 2.50 & 2.81 & 2.01 & 61.1 & 78.8 & 87.3 & 91.2 & 94.9 & 82.7  \\
    & FRCRN & 6.90M & 1.26 & 1.67 & 2.28 & 2.74 & 3.12 & 2.21 & 70.8 & 84.4 & 90.9 & 94.4 & 96.7 & 87.4  \\
    & CMGAN & 1.92M & 1.37 & 1.77 & 2.41 & 2.89 & 3.28 & 2.34 & 76.9 & 86.1 & 92.0 & 95.1 & 97.2 & 89.5  \\
    & MP-SENet & 2.05M & 1.45 & 1.93 & 2.54 & 3.01 & 3.39 & 2.46 & 78.9 & 87.5 & \textbf{92.9} & 95.7 & \textbf{97.5} & 90.5  \\
    & MUSE & 0.51M & 1.24 & 1.64 & 2.23 & 2.70 & 3.06 & 2.17 & 68.2 & 83.9 & 90.6 & 94.1 & 96.6 & 86.7  \\
    & MP-SENet (S) & 1.16M & 1.38 & 1.79 & 2.43 & 2.92 & 3.31 & 2.37 & 77.2 & 86.3 & 92.1 & 95.2 & 97.2 & 89.6  \\
    & MUSE (L) & 1.14M & 1.29 & 1.70 & 2.32 & 2.79 & 3.17 & 2.25 & 73.1 & 84.9 & 91.3 & 94.6 & 96.8 & 88.1  \\
    \hline
    & \textbf{LORT} & 0.96M & \textbf{1.58} & \textbf{2.03} & \textbf{2.63} & \textbf{3.08} & \textbf{3.45} & \textbf{2.55} & \textbf{79.1} & \textbf{87.6} & \textbf{92.9} & \textbf{95.8} & \textbf{97.5} & \textbf{90.6}  \\
    \hline
  \end{tabular}
  \label{tabDNS2}
\end{table*}

The results clearly demonstrate that LORT achieves highly competitive performance across all objective evaluation metrics. With only 0.96M parameters and 16.83G FLOPs, LORT achieves a PESQ of 3.51, CSIG of 4.74, and COVL of 4.23—surpassing all models with fewer than 1 million parameters, including the previously best-performing MUSE (0.51M, 9.43G) and S4ND U-Net (0.75M). Remarkably, LORT even surpasses several SOTA models with significantly higher model complexity. For example, compared to MP-SENet, which has more than double the parameters and over four times the computational cost (FLOPs), LORT achieves better performance on all metrics except CBAK and STOI, indicating its superior efficiency in both architectural design and representation capacity.

To further validate the robustness of LORT under similar resource constraints, we perform a controlled comparison by adapting two strong baseline models, MP-SENet and MUSE, to match the parameter scale of LORT. Specifically, for MP-SENet, we reduce the number of input channels from 64 to 48, and for MUSE, we increase the number of input channels from 16 to 22. These adapted variants, denoted as MP-SENet (S) and MUSE (L) respectively, contain approximately 1M parameters. As shown in Table~\ref{tab4}, LORT significantly outperforms both variants across all evaluation metrics, despite using fewer FLOPs than MP-SENet (S) and having nearly the same FLOPs as MUSE (L). This result further confirms the effectiveness and scalability of the LORT architecture, demonstrating its potential for real-world, resource-constrained applications.

\subsection{Comparison with Baselines on DNS Challenge dataset}

Table \ref{tabDNS2} presents a comprehensive performance comparison of LORT with several baseline models across three different noise conditions: DNS Challenge dataset noise, Babble noise, and Factory noise. All models are evaluated across five SNR levels ranging from -5 dB to 15 dB, using PESQ and STOI as the primary evaluation metrics to reflect perceptual speech quality and intelligibility.

\subsubsection{Performance under DNS Challenge noise}

In the general DNS noise scenario, LORT demonstrates dominant performance across all SNR levels. For PESQ, LORT achieves scores of 2.26 (-5 dB), 2.75 (0 dB), 3.14 (5 dB), 3.53 (10 dB), and 3.92 (15 dB), with an average of 3.12—surpassing the second-best MP-SENet (3.04) and other baselines. This superiority is particularly notable at low SNR (-5 dB), where LORT outperforms MP-SENet by 0.11 points, indicating strong noise suppression in severely degraded conditions. In terms of STOI, LORT matches MP-SENet with an average of 95.0\%, achieving 89.0\% at -5 dB and reaching near-perfect intelligibility (98.6\%) at 15 dB. When compared to parameter-matched baselines—MP-SENet (S) (1.16M parameters) and MUSE (L) (1.14M parameters)—LORT (0.96M parameters) shows significant advantages: +0.20 higher average PESQ than MP-SENet (S) and +0.43 over MUSE (L), alongside +0.5\% and +2.5\% improvements in average STOI, respectively. This highlights LORT’s efficiency in balancing model parameter and performance.

\subsubsection{Performance under Babble and Factory noise}

Under non-stationary Babble noise, LORT consistently outperforms all baselines in both metrics. The PESQ scores range from 1.37 (-5 dB) to 3.51 (15 dB), averaging 2.52—0.12 higher than MP-SENet (2.40) and 0.47 above MUSE (2.05). At the lowest SNR, LORT’s PESQ is 0.10 higher than MP-SENet, demonstrating superior handling of complex, interfering speech patterns. For STOI, LORT achieves an average of 89.3\%, matching MP-SENet’s performance while outperforming other baselines across all SNRs. Compared to adjusted baselines, LORT surpasses MP-SENet (S) and MUSE (L) by +0.25 and +0.26 in average PESQ, and +1.0\% and +3.4\% in average STOI, respectively, indicating its adaptability to non-stationary noise environments.

In the Factory noise scenario, LORT performs outstandingly again, achieving the highest average PESQ (2.55) and STOI (90.6\%). Its PESQ score ranges from 1.58 at -5 dB to 3.45 at 15 dB, with an average of 2.55, significantly outperforming MP-SENet (with an average of 2.46) and MUSE (with an average of 2.17). For STOI, LORT attains the best performance, reaching 79.1\% at -5 dB and 97.5\% at 15 dB, which is comparable to the best baseline performance. The average STOI of LORT is 90.6\%, 0.1\% higher than that of MP-SENet and 3.9\% higher than that of MUSE. When compared with the adjusted SOTA baselines with similar parameters, the average PESQ of LORT is 0.18 and 0.30 higher than those of MP-SENet (S) and MUSE (L) respectively, and its average STOI is 1.0\% and +2.5\% higher than those of MP-SENet (S) and MUSE (L) respectively. These results signify the effectiveness of LORT in dealing with Factory noise.

Overall, across all three types of noise, LORT consistently exhibits superiority in both perceptual quality and intelligibility across different SNRs, whether compared to larger models such as MP-SENet and CMGAN, as well as to baselines with similar parameter counts, including MUSE (L) and MP-SENet (S). These findings highlight the efficiency and robustness of LORT in speech enhancement.

\section{Conclusion and future work}
\label{conclusions}

This study presents a lightweight and efficient speech enhancement model, LORT, which achieves remarkable enhancement performance while ensuring low computational overhead. The proposed architecture is designed to comprehensively learn both global and local speech information. Specifically, the novel Taylor multi-head self-attention (T-MSA) module incorporating spatial-channel enhancement attention (SCEA) is primarily used for coarse-grained global modeling, which not only maintains linear computational complexity but also effectively facilitates inter-channel communication while mitigating the inherent spatial attention limitations of Taylor-based Transformers. For local modeling, the locally refined convolution (LRC) block is designed to effectively capture fine-grained local speech features via the combination of convolutional feedforward networks, time-frequency dense convolutions, and gating units, thereby compensating for the limitations of Taylor Transformer in modeling local dependencies. Implemented within a U-Net-style encoder-decoder framework with alternating downsampling and upsampling paths, LORT efficiently learns multi-resolution speech representations while requiring only 0.96M parameters and 16.83G FLOPs of computational complexity. Experiments on the VCTK +DEMAND and DNS Challenge datasets show that LORT can reach or even surpass the performance of current mainstream state-of-the-art (SOTA) models while significantly reducing the model size and computational amount, indicating its great application potential in complex acoustic and resource-constrained scenarios.

In the future, we will further explore the generalization ability of LORT in other speech processing tasks such as dereverberation and multi-speaker enhancement, and strive to enhance its robustness and adaptability under low-SNR or cross-domain acoustic conditions.

 \bibliographystyle{elsarticle-harv} 
\bibliography{ref}

\end{document}